# Detection of Leonids meteoric dust in the upper atmosphere by polarization measurements of the twilight sky

Oleg S. Ugolnikov[a,b,*], Igor A. Maslov[a,c]

[a]Space Research Institute, Profsoyuznaya st., 84/32, 117997, Moscow, Russia
[b]Astro-Space Center, Lebedev's Physical Institute, Profsoyuznaya st., 84/32, 117997, Moscow, Russia
[c]Sternberg Astronomical Institute, Universitetsky prosp., 13, 119992, Moscow, Russia

*Corresponding author. Tel.: +7 495 333 4011; Fax: +7 495 333 5178
E-mail address: ugol@tanatos.asc.rssi.ru (O.S. Ugolnikov), imaslov@iki.rssi.ru (I.A. Maslov)

**Abstract.** The method of detection of dust in the stratosphere and mesosphere by the twilight sky background observations is being considered. The polarization measurements are effective for detection of the meteoric dust scattering on the background consisting basically of troposphere multiple scattering. The method is based on the observed and explained polarization properties of the sky background during different stages of twilight. It is used to detect the mesosphere dust after the Leonids maximum in 2002 and to investigate its evolution. The polarization method takes into account the multiple scattering and sufficient contribution of moonlight scattering background and turns out to be more sensitive than existing analogs used in the present time.

## 1. Introduction.

Wide possibilities of atmosphere sounding by twilight background photometry are known for a long time. If we measure the sky background intensity during the day, we will obtain the integral characteristics of the air column above the observer, those are contributed basically by lower dense layers of the atmosphere (Fesenkov, 1955). Upper layers do not influence on the daytime background. But as the twilight comes, the Earth's shadow covers much higher atmosphere layers, and the effective altitude of single scattering rises with the Sun depression under horizon. Having measured the twilight sky intensity, we obtain the information about definite layer of the atmosphere. Continuing the measurements during the whole twilight period, we build the vertical atmosphere scan. This is the basic principle of twilight sounding method, being developed from the pioneer work (Fesenkov, 1923).

Optical properties of stratosphere and mesosphere depend on the dust and aerosol density in these layers and can sufficiently change after the maxima of major meteor showers, which was observed by (Link and Robley, 1971, Link, 1975) as increase of twilight sky background intensity during different stages of twilight. The point of the sky and solar zenith angle when it occurs are corresponding to definite altitude of Earth's shadow and effective altitude of scattering. This becomes the base of the method of detection of meteoric and volcanic dust in the atmosphere. The method was used to detect the dust from a number of major showers, such as Quadrantids (Link and Robley, 1971), Orionids (Mateshvili and Mateshvili, 1988) and η-Aquarids (Mateshvili et al., 1997) during the return of parent comet Halley, and others. In late 1990s and early 2000s the observations were basically focused on the return of Leonids, which occurs once in 33 years producing the burst of activity up to several thousands and dozens of thousands meteors per hour. The twilight sky background increase was detected during the Leonids burst period in 1998 (Mateshvili et al., 1999) and 1999 (Mateshvili et al., 2000).

Method used in papers above was based on the assumption that all observational properties of the twilight sky are related with the atmosphere layers emitted by straight solar radiation in this moment. Actually it is the single scattering model. Being working well during the daytime period, this model meets serious problems during the twilight, when the solar emission is transferred along optically thick tangent path in the lower atmosphere layers. Possible sufficient contribution of multiple scattering in the twilight background was noticed in (Fesenkov, 1923) and became the subject of detailed analysis in (Rozenberg, 1966, see also references therein). Although the estimation of multiple scattering contribution during the light twilight period was principally correct, it was sufficiently underestimated later, during the dark twilight period with solar zenith angle SZA>95°, when multiple scattering starts to take over the single scattering.

The contribution of multiple scattering during different twilight stages was estimated by polarization observations in (Ugolnikov, 1999) for the wavelength 356 nm, than, in (Ugolnikov and Maslov, 2002) the same was done for the number of bands in the visible part of spectrum. As it was shown, the multiple scattering contribution near the sunrise (or sunset) rises from about 30% in the red spectral region up to 60% in violet one in the case of stable clear skies. Being almost constant until the solar zenith angle 94-95°, the multiple scattering contribution increases after that, practically reaching 100% at the solar zenith angle 98-99°. These results were confirmed by the numerical solution of radiative transfer equation made by vector code MCC++ (Postylyakov, 2004). Correct numerical account of multiple scattering, which became available recently, leads to satisfactory agreement of twilight background intensity and polarization calculations with the observational data (Ugolnikov et al., 2004, Patat et al., 2006). The total domination of multiple scattering in the dark period of twilight (solar zenith angle about 99-100°) can be also proved by simple observational analysis (see the Chapter 3).

If the meteoric dust appears in the upper atmosphere, the corresponding scattering fraction will be observed on the background of multiple scattering forming in the other layers of the atmosphere. The intensity of this background depends on the transparency and scattering properties of these lower atmospheric layers not only at the observation place, but also along the whole trajectory of photons diffusion. Such dependence is quite complicated, and the clear sky intensity changes from twilight to twilight. It is the basic problem during the separation of single and multiple scattering based on the intensity measurements only. Detection of meteoric dust and estimation of particle density based on intensity comparison for different twilight periods with single scattering model may lead to systematical errors.

More exact approach to this problem was suggested in (Mateshvili and Rietmeijer, 2002) and used in (Padma Kumari et al., 2005) for the analysis of Leonids meteoric dust in early 2000s. The basic observational parameter there is not the twilight background intensity, but its logarithmic derivative by the effective altitude of single scattering. Rapid variations of such value during the twilight are interpreted as the influence of meteoric dust layer at the corresponding altitude. It is correct since the multiple scattered background cannot have short-period variations during the twilight. The method is effective for detection of dense meteoric dust layers, but the numerical estimations need the multiple scattering properties to be taken into account, that is quite complicated problem.

The primary goal of this work is to develop the method of detection of meteoric dust in the atmosphere based on the polarization measurements of the twilight sky. Since the polarization properties sufficiently differ for various twilight sky components (single Rayleigh scattering, single aerosol scattering, multiple scattering), polarization measurements are the independent and effective tool for their separation, as it was already noticed in (Fesenkov, 1966). Polarization method was used to divide single and multiple scattering (Ugolnikov, 1999, Ugolnikov and Maslov, 2002), to detect and investigate the aerosol scattering in the troposphere (Ugolnikov and Maslov, 2005b). It is quite sensitive, being able to detect weak optical features in the atmosphere, invisible for the intensity probes. It can work in the case of moonlit twilight. In this work this method will be applied to the Leonids meteoric dust layer investigation in 2002, the last year of strong Leonids burst after the 1998 perihelion of parent comet, 55P/Tempel-Tuttle.

## 2. Observations and dust fall conditions.

Polarization observations of twilight sky background were conducted in Crimean Laboratory of Sternberg Astronomical Institute (Crimea, Ukraine, 44.7°N, 34.0°E, 600 m a.s.l.). The observational device was the wide-angle CCD-camera (8°x6°) directed to the zenith with rotating polarization filter. The data were averaged over the square 1°x1°. The observations were carried out in the wide spectral band with effective wavelength 525 nm. This band is close to Johnson-Cousins V band, for the stars magnitudes in nighttime images we have (m–V)=0.06·(B–V). The polarization filter is practically ideal (better than 99.9%) in this spectral band. The measurements started during the day, when the Sun was still above the horizon (solar zenith angle about 87°), and continued deep into the night, and then, till the same solar zenith angle in the morning. The basic data were obtained during the Leonids activity epoch in 2002.

Investigating the meteoric dust inflow in the definite location on the Earth, it is necessary to check its conditions in this location for the shower maximum being considered, this check is unfortunately missing in a number of papers. The maximum time and activity of Leonids in 2002 were successfully predicted by Lyytinen and van Flandern (2000), McNaught and Asher (2002), Vaubaillon (2002). Two strong activity peaks were registered by observations (Arlt et al., 2002). First one was observed during the Earth encounter with 1767 dust trail of comet 55P/Temple-Tuttle, which had experienced a number of the same events during its 7 orbital revolutions. The last one in 2001 had provided unexpectedly large Zenithal Hourly Rate value (about 1600, McNaught and Asher, 2002). In 2002, the activity peak took place at $04^h10^m$ UT, November, 19. Zenithal Hourly Rate reached 2500, the full width at half-maximum was about 40 minutes (Arlt et al., 2002). During this maximum the Leonids radiant was at 67° above the horizon at the observation place, providing good conditions for the dust inflow to the upper atmosphere. Second Leonids maximum occurred during the encounter with 1866 dust trail at $10^h50^m$ UT in the same day. The radiant altitude in the observation place was about 8°, and in spite of large Zenithal Hourly Rate (up to 3000), the value of dust inflow in the observation place is small compared with the first peak.

Polarization observations of the zenith twilight sky had started on November, 21, when evening twilight was covered by measurements. The basic problem of the Leonids investigation in 2002 was the full Moon near the maximum, causing the difficulties during meteors registration and sufficiently emitting the twilight sky. In the evening of November, 21, moonrise occurred during the twilight, when solar zenith angle was equal to 100°. Next observations were conducted in the morning and evening of November, 27, and the last observation of this month was carried out in the morning of November, 30. Both morning twilights of November, 27 and 30, were also enlighten by waning Moon.

Results of twilight measurements in post-maximum epoch of Leonids in 2002 were compared with moonless data obtained in the same location with the same spectral band outside this period. The morning twilight data of December, 11, 2002, can be used as the basic reference twilight data, since the maximal atmosphere transparency and the twilight sky polarization were registered alongside with vanishing level of tropospheric aerosol (Ugolnikov and Maslov, 2005a, 2005b). Other twilight data of December 2002 and different seasons of 2000 and 2006 were also used for comparison. It must be noted that the time amount after the last sufficient volcanic eruption (Mount Pinatubo in 1991) is quite large and the stratosphere was clear from aerosol, as it was shown in (Ugolnikov and Maslov, 2002, 2005b).

## 3. Twilight background polarization properties.

In this chapter the dependency of the twilight background polarization in the zenith point on the solar zenith angle is considered in the case of clear skies, lack of stratospheric and mesospheric aerosol and low amount of tropospheric aerosol. The twilight background near the zenith is strongly polarized due to the almost linear polarization of the Rayleigh scattering by the angle 90°, the value reaches 0.94 for the atmospheric air. Real polarization value is lower due to several depolarization factors (multiple scattering, aerosol scattering, moonlight scattering). Let us start considering the simplest case.

*3.1. Moonless conditions.*

It is the case used as criteria for the basic reference twilight data in this work. The twilight sky background is forming by the scattering of solar emission (of different order) until the solar zenith angle about 101°, when the less polarized night sky background (Ugolnikov and Maslov, 2005a) starts to make remarkable contribution. Let's build the horizontal reference frame connected with the Sun (where solar azimuth is equal to zero). From the symmetry of the picture, the three-dimension Stokes vector of the twilight background $S_0$ in the zenith point takes the following form:

$$S_0 = \left\{ \begin{array}{c} I_0(SZA) \\ -I_0(SZA) \cdot p_0(SZA) \\ 0 \end{array} \right\} \qquad (1),$$

where $I_0$ and $p_0$ mean the brightness and polarization of the twilight sky, depending on the solar zenith angle.

Figure 1 shows the dependency of zenith twilight sky polarization on the solar zenith angle for a sample of moonless twilights in the instrumental spectral band. The same behavior of such dependency for all twilight in different years and seasons is clear to see. The twilight period can be separated into different stages:

1. *Early twilight*, from daytime till SZA about 90-91°: the polarization slowly rises with SZA due to tropospheric aerosol immersion into the effective shadow of the Earth and vanishing of aerosol depolarization. The altitude of this shadow is higher than the geometrical one due to strong absorption of tangent emission (Ugolnikov, 1998), being about 15 km at the sunset moment for observational spectral band and slightly rising after that. The effect of polarization change is almost absent during the basic reference morning twilight of December, 11, 2002, when the amount of tropospheric aerosol is minimal.

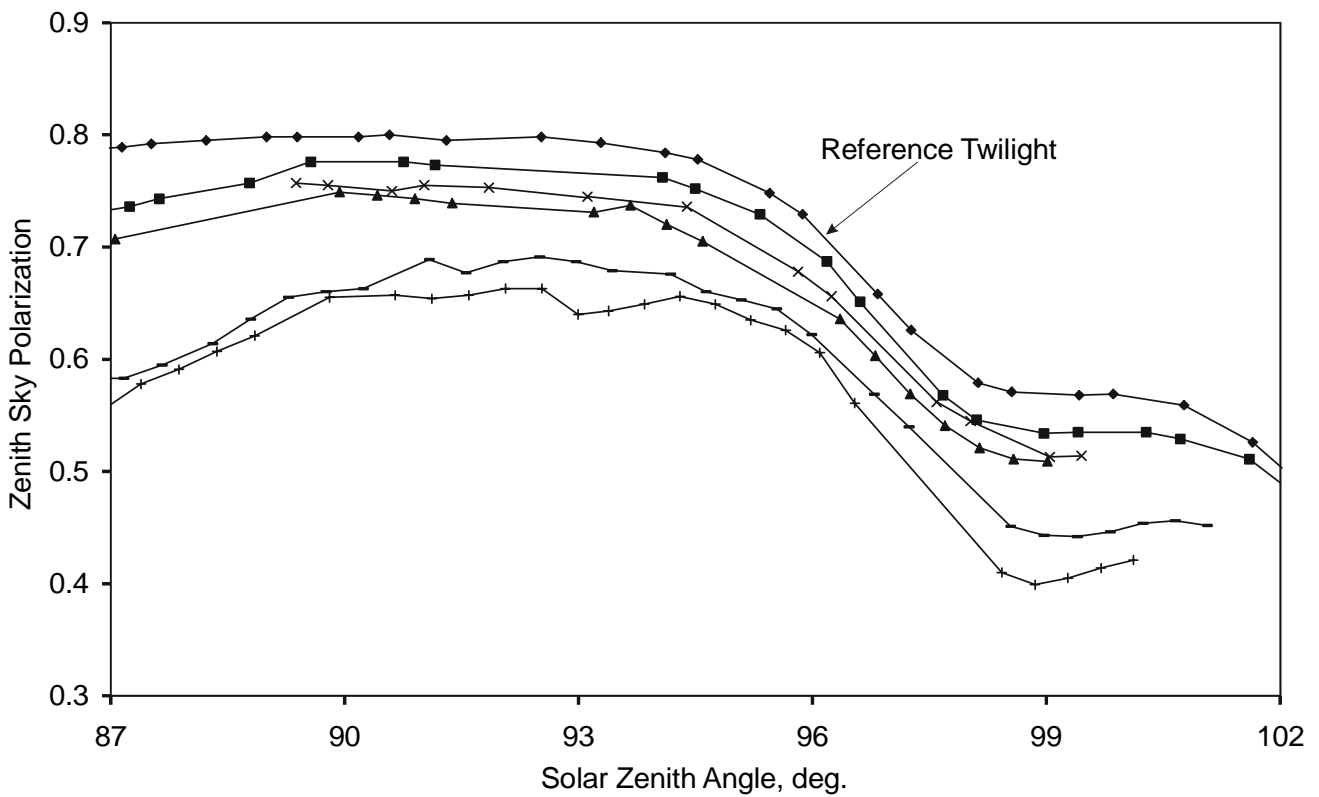

Figure 1. The dependency of twilight sky background polarization in the zenith on the solar zenith angle for the sample of the moonless observations in July-August 2000 and November-December 2002. The dependency for the reference twilight of the morning, December, 11, is marked.

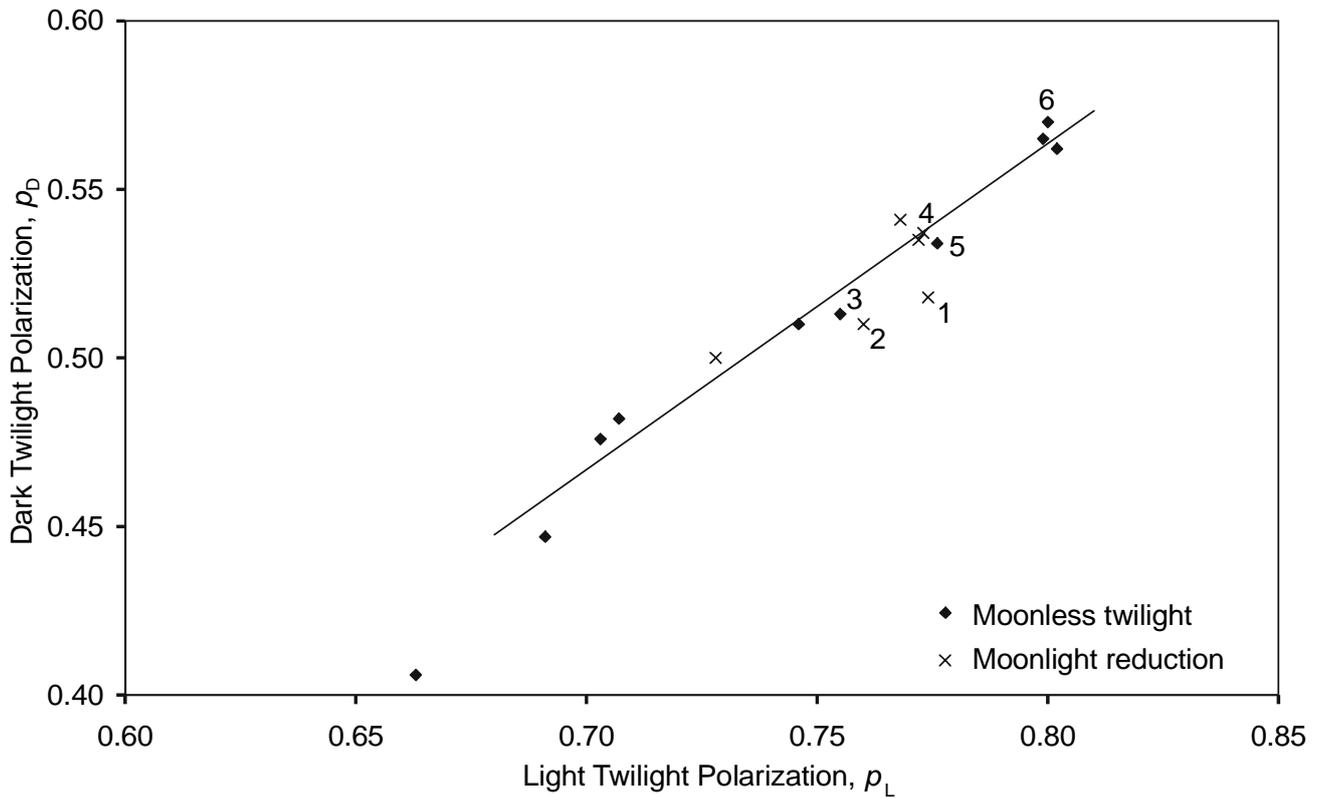

Figure 2. Correlation of zenith sky polarization values during the light and dark stages of twilight for July-August 2000, November-December 2002 and October 2006 (1 – evening, November, 21; 2 – morning, November, 27; 3 – evening, November, 27; 4 – morning, November, 30; 5 – morning, December, 9; 6 – reference twilight, morning, December, 11, 2002).

2. *Light twilight*, SZA from 90-91° to 94-95°: the polarization in the zenith reaches the maximum and then is almost constant, starting to decrease in the end of this period. The background consists of single Rayleigh scattering and troposphere multiple scattering. Their ratio remains practically constant and can be estimated by method suggested in (Ugolnikov, 1999, Ugolnikov and Maslov, 2002). The intensity contribution of single scattering for the instrumental spectral band is about 0.6 (0.59 in the evening of November, 21, 0.60 and 0.57 in the morning and evening of November, 27, respectively, 0.64 for the morning of November, 30 and for the basic reference morning twilight of December, 11).

3. *Transitive twilight*, SZA from 94-95° to 98.5°: fast decrease of the sky polarization, caused by fading of single scattering component on the background of multiple scattering. This also changes the sky color and brightness distribution. It is the period of fastest decrease of sky brightness, that's why the "civil twilight" is assumed to fade at SZA equal to 96°.

4. *Dark twilight*, SZA from 98.5 to 100.5°: the sky polarization is constant again, being close to the polarization of multiple scattering background component during the light twilight (Ugolnikov and Maslov, 2005b). The twilight sky background near the zenith completely consists of multiple scattering.

5. *Nightfall*, SZA from 100.5° to 108°: polarization decreases and then vanishes due to increasing contribution of night sky background.

Of course, this classification depends on the wavelength. For example, in violet part of spectrum the light twilight starts before the sunset, and in the near infrared bands there is no light and dark twilight, early twilight turns to transitive and then to nightfall. Classification described above for the best visible part of spectrum makes this band suitable for the upper meteoric dust detection. Appearing at the altitudes about 80-100 km, this dust changes the constant polarization of dark twilight background, which can be easily detected during the observations.

As we can see in the Figure 1, the dependencies of polarization on the solar zenith angle are principally the same, being just shifted one from another. This fact can be also illustrated by the diagram in the Figure 2, where the comparison of maximal sky polarization during the light twilight $p_L$ and the "plateau" polarization during the dark twilight $p_D$ is shown for the moonless twilight data (filled points). We see the excellent correlation. Since the first of these values, $p_L$, is determined by optical conditions of the lower atmosphere, the correlated value, $p_D$, must be determined by the same atmospheric layers, that are immersed into the shadow of the Earth during the dark twilight. Thus, the polarization of the dark twilight sky at SZA near 99-100° is determined by multiple scattering, and here is the simplest way to show it.

Day-to-day variability of the twilight sky polarization dependencies is related with the ground-based aerosol, changing the brightness and polarization of multiple scattering and shifting the curve in the Figure 1 as a whole. The Figure 2 shows that the correlation of $p_L$ and $p_D$ is linear for $p_L>0.7$, that is true for the most part of observations, the linear relation coefficient is close to unity. It is the property of the twilight sky in yellow-green part of spectrum that will help us to build the meteoric dust detection method.

*3.2. Moonlit twilight.*

This case is more complicated than the previous one. Scattered moonlight may be disregarded during the early and light twilight, since its contribution in the sky background is unnoticeable. It remains quite small during the transitive twilight, but starts to dominate during the dark phase. In fact, there is no nightfall period in this case, the dark twilight turns to the "lunar day" or "lunar twilight" depending on the position of the Moon in the sky. The sky polarization becomes similar to the daytime value for the same altitude of the Sun above the horizon. The Stokes vector of the sky background in the zenith in the same coordinate system takes the following form:

$$S = S_0 + S_L = \begin{Bmatrix} I_0(SZA) + I_L(LZA) \\ -I_0(SZA) \cdot p_0(SZA) - I_L(LZA) \cdot p_L(LZA) \cdot \cos 2\Delta A \\ -I_L(LZA) \cdot p_L(LZA) \cdot \sin 2\Delta A \end{Bmatrix} \quad (2),$$

where $I_L$ and $p_L$ are the brightness and polarization of scattered moonlight depending on the lunar zenith angle LZA, $\Delta A$ is the azimuth difference of the Moon and the Sun. The resulting polarization will depend on the brightness and position of the Moon in the sky. When it is close to the first or last quarter and $\Delta A$ is about 90°, the moonlight scattering will decrease the polarization of the twilight sky during the dark twilight period (see the dependency for the morning, November, 27, in the Figure 3). If the Moon is close to full, the polarization can either increase or decrease, depending on the value of $p_L$(LZA).

If Moon is high above the horizon during the whole twilight period and the beginning (or the end) of night and its altitude is not changing fast, than it is possible to measure the dependencies of $I_L$ and $p_L$ on the lunar zenith angle during the night. After the reduction the night sky background this dependency can be extrapolated to the twilight period. It gives us the possibility to subtract the vector $S_L$ from the measured data. This procedure can be done for the moonlit twilights, the result for the morning of November, 27 is shown in the Figure 3 by the dashed line. Since the night sky background is estimating for the nearby moonless period, this procedure can be run when the influence of night sky background variations can be neglected. It is true for SZA<102° during the moonlit twilight. Moonlight corrected data for the sample of twilights are shown by the cross symbols in the Figure 2.

But the procedure doesn't work if the Moon phase is close to full and situated near the horizon during the twilight period. In this case of "lunar twilight" the dependencies $I_L$ and $p_L$ become rapid and can not be extrapolated. This situation took place in the evening of November, 21, 2002, when the Moon rose at SZA equal to 100°. This case the values of $I_L$ and $p_L$ can be calculated using the following assumption:

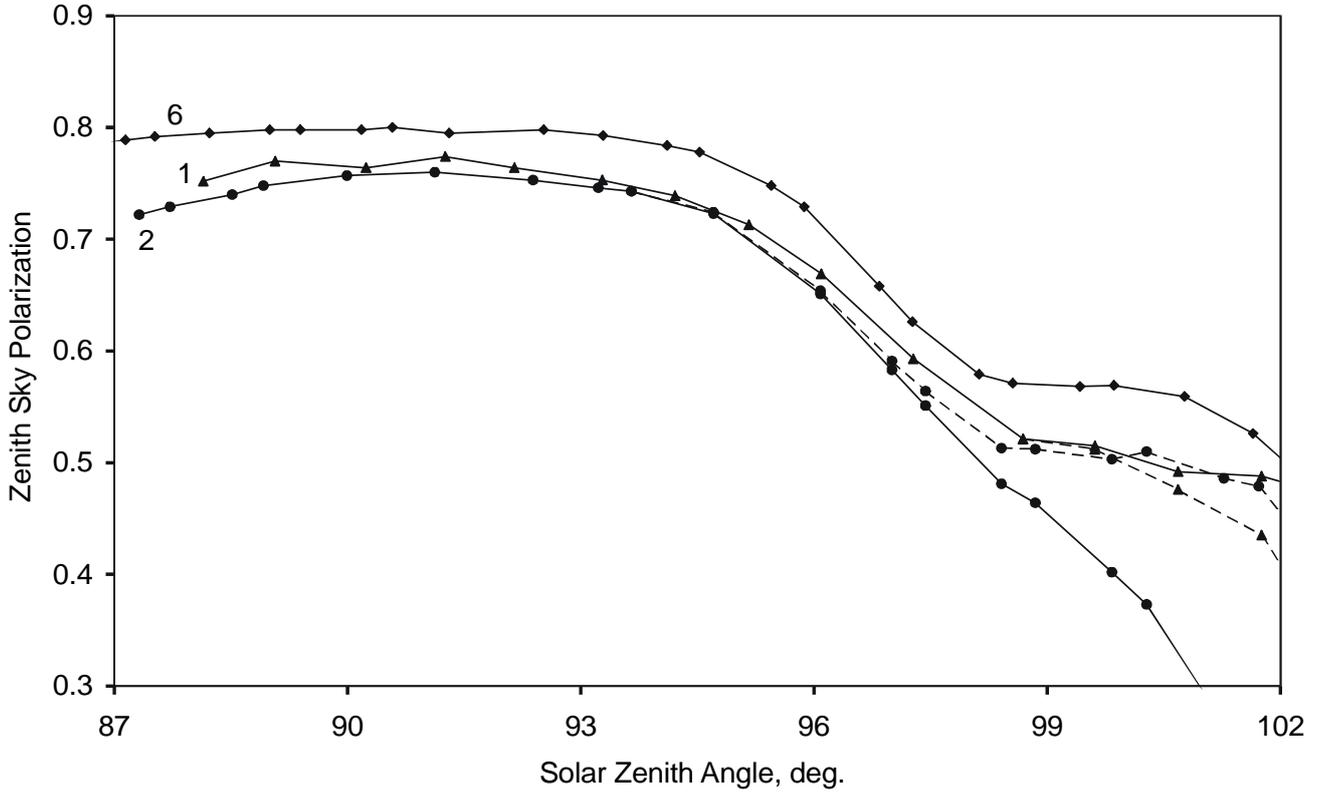

Figure 3. The dependency of twilight sky background polarization in the zenith on the solar zenith angle for moonlit twilights of evening, November, 21 and morning, November, 27, 2002, compared with the one for reference twilight. Digital designations are analogous to the Figure 2, dashed lines show the moonlight reduction results.

$$\frac{I_L(z)}{I_0(z)} = \frac{J_L}{J_0}; \quad p_L(z) = p_0(z) \tag{3},$$

where $J_0$ and $J_L$ are the brightness values for the Sun and the Moon for the observation time, $z$ is the zenith distance of Sun or Moon. Actually it is the assumption of atmosphere symmetry and stability during the observations. It can be used when the lunar emission scattering is not dominating in the sky background, that is true again for SZA<102° in the evening of November, 21. The result of moonlight reduction for this evening is also shown by cross symbol in the Figure 2 and the dashed line in the Figure 3. We see that full moonlight scattering near the horizon is increasing the twilight sky polarization at the zenith during the dark twilight and nightfall. It is quite natural since the Sun and the Moon are situated in almost opposite directions, and the scattered moonlight is sufficiently polarized when the Moon is close to the horizon.

*3.3. Influence of meteoric aerosol.*

As it can be seen from the observation data, the dark twilight sky polarization in the zenith is higher than 50%. It is the property of multiple scattering, that also remains during the light and transitive twilight. The polarization of light scattered on the dust particles (including the ones with submicron sizes) for the considered spectral band is less than 50%. Thus, the contribution of meteoric dust scattering should decrease the total polarization of the twilight sky independently on the polarization angles and brightness ratio of different background components. The depolarization effect will be detected, when the effective altitude of single scattering corresponds to the dust layer. The same effect caused by tropospheric aerosol is observed during the early twilight (Ugolnikov and Maslov, 2005b). The effective altitude of single scattering is related with the solar zenith angle by the following equation:

$$H(SZA) = \frac{R+H_0}{\sin(SZA)} - R \approx H_0 + \frac{R \cdot (SZA - (\pi/2))^2}{2} \qquad (4).$$

Here SZA is expressed in radians, $R$ is the radius of Earth and $H_0$ is the optical excess of Earth's shadow caused by absorption of solar tangent emission. This value (unfortunately missed in some papers) is almost constant for SZA>94°, being equal to 18.8 km for our spectral band.

**4. Detection of meteoric dust light scattering.**

Let us return to the Figures 1 and 2. As we saw there, in the moonless conditions (or if the moonlight scattering is subtracted) in the absence of dust and aerosol, the polarization values $p_L$ and $p_D$ during the light and dark twilight are correlated, and if maximal twilight polarization $p_L$>0.7 (good atmosphere conditions), this relation is linear with coefficient close to unity. So, the difference $(p_D - p_L)$ is principally the same for all twilights. Figure 1 shows that it is true not only for dark twilight, but for transitive twilight and start of nightfall. The moonless difference $(p_0(SZA) - p_L)$ is the function of SZA, being the same for all twilights too. Let us subtract from this difference the one interpolated for the basic reference twilight, the morning of December, 11, 2002:

$$\Delta p(SZA) = (p_0(SZA) - p_L) - (p_0(SZA) - p_L)_{ref} \qquad (5).$$

In the absence of dust and aerosol this value is expected to be close to zero, the meteoric dust contribution must lead to the negative values of $\Delta p$ (decrease of polarization $p_0$) for definite interval of SZA, according to the formula (4).

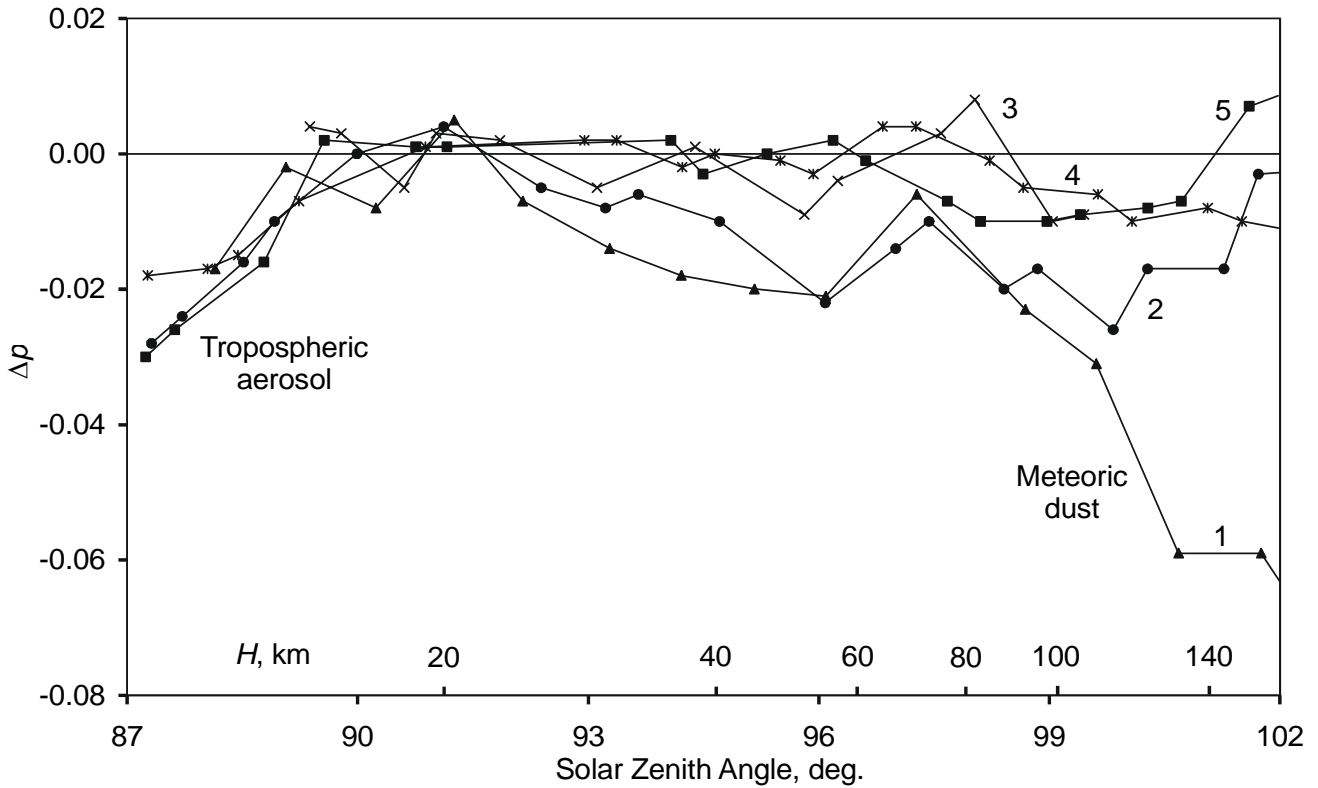

Figure 4. The polarization change quantity $\Delta p$ as the function of solar zenith angle and effective scattering altitude. Digital designations are analogous to the Figure 2.

Figure 4 shows the dependencies Δ$p$(SZA) for different twilights in November and December, 2002, other than the reference one. The values of effective scattering altitude are also shown at the x-axis. We see that this value of Δ$p$ is really close to zero for the most part of cases. The depolarization appears in some cases during the early twilight, this effect is caused by tropospheric aerosol and was investigated in (Ugolnikov and Maslov, 2005b). Depolarization also appears during the dark stage of evening twilight at November, 21, first one in the observations set in 2002. This effect also leads to deviation of $p_D$ value for this evening in the Figure 2 and can be seen in the Figure 3 as the lack of polarization. Since this Δ$p$ value is not reached during any other twilight, it is quite natural to relate it with Leonid meteoric dust at the altitudes above 90 km. However, the depolarization is also present at lower altitudes.

Depolarization effect is seen for the next twilight been observed, the morning of November, 27, 2002, with less value and wide distribution by the altitudes in the stratosphere and mesosphere. The corresponding point in the Figure 2 is also shifted downwards. This effect is almost disappearing during the evening twilight of the same day, and after that the module of Δ$p$ does not exceed the background value of 0.01 for all twilight stages, except the early twilight. Due to the low frequency of good weather conditions, just two twilights with noticeable influence of Leonid meteoric dust were observed. In spite of this fact and contribution of moonlight scattering during both twilights, the dynamics of dust layer, appearing in the mesosphere and lower thermosphere and than spreading down to the stratosphere, is clearly seen.

## 5. Discussion and conclusion.

In this paper the polarization method of meteoric dust detection basing on the twilight observations is considered. Using this method, the depolarization of the twilight sky in the zenith during the dark twilight stage was found in the evening of November, 21, 2002, 2.5 days after the strong Leonid maximum caused by comet 55P/Temple-Tuttle 7-revolution dust trail. During this maximum one hemisphere of the Earth, including the observation point, was bombarded by meteoric dust. The values of solar zenith angle, where the maximal depolarization occurs, correspond to the altitudes higher than 90 km. Weaker depolarization is noticeable for the next observed twilight, the morning, November, 27, covering the wide range of solar zenith angles and effective scattering altitudes in stratosphere and mesosphere. This effect is unseen during the other twilights of different years and seasons, when observations were made, that makes the relation of depolarization with meteoric dust quite obvious.

Polarization method turns out to be more sensitive than intensity tests used in a number of papers (see the Chapter 1). This fact is illustrated in the Figures 5 and 6, containing the altitude dependencies of sky brightness ratio of given and reference twilight and the logarithmic derivative of brightness by the effective scattering altitude. These two values are used in (Mateshvili et al., 1999, 2000) and (Padma Kumari et al., 2005), respectively, to detect the meteoric dust scattering component. We see that all the dependencies are smooth, being close to each other and have no noticeable features. The sky brightness is usually less than the one for reference twilight due to less atmospheric transparency, it is the effect of multiple scattering as it was noted in Chapter 1. Brightness increase is seen for the evening, November, 21, at altitudes higher than 130 km, but it is model-dependent due to sufficient contribution of reduced moonlight scattering during the late dark twilight and nightfall period.

We see that if the meteoric dust layer expand over wide range of altitudes, and its contribution to the twilight background intensity does not exceed 10-20%, it will not cause noticeable changes of brightness and its logarithmic gradient, but anyway it can be detected by polarization measurements. The same is true for the light scattering contribution of tropospheric aerosol during the early twilight (Ugolnikov and Maslov, 2005b). Polarization method also takes into account the domination of multiple scattering during the dark stage of twilight, when meteoric dust scattering can be observed.

The meteoric dust layer is found to be thick, no stratification is observed. The possible reason is high altitude of Leonids radiant above the horizon at the observation place during the first maximum in 2002, allowing the meteoric dust to inflow at wide range of altitudes. The meteoric dust layer at November, 27, 2002 is higher than it is reported in (Padma Kumari, 2005) for the same day. It can be related with different value of optical excess of the Earth's shadow $H_0$ taken in that paper and better sensitivity of polarization method at higher altitudes.

Yellow-green part of spectrum turns out to be the best range for meteoric dust investigations. From the one side, the intensity ratio of dust scattering and troposphere multiple scattering is sufficiently larger than at shorter wavelengths. From the other side, the quantity of $H_0$ is higher than at longer wavelengths, the troposphere aerosol is already immersed into the shadow during the sunset, and we have the definite period of light twilight clear from aerosol depolarization (owing to the absence of volcanic dust), which is important for the method considered.

The polarization method of meteoric dust detection can be improved by wide-angle data, giving the possibility to measure the vector scattering function of meteoric dust and investigate its microphysical characteristics. It will also give more information about the altitude distribution of meteoric dust. Since the sky brightness during the dark twilight is basically contributed by troposphere multiple scattering and depends on the altitude of the observation place (Patat et al., 2006), the observations are better to conduct in the mountain sites, where this background is lower.

**Acknowledgements.** Authors are thankful to N.N. Shakhvorostova (Astro-Space Center of Lebedev's Physical Institute) for the useful remarks. O.S. Ugolnikov is supported by Russian Science Support Foundation grant.

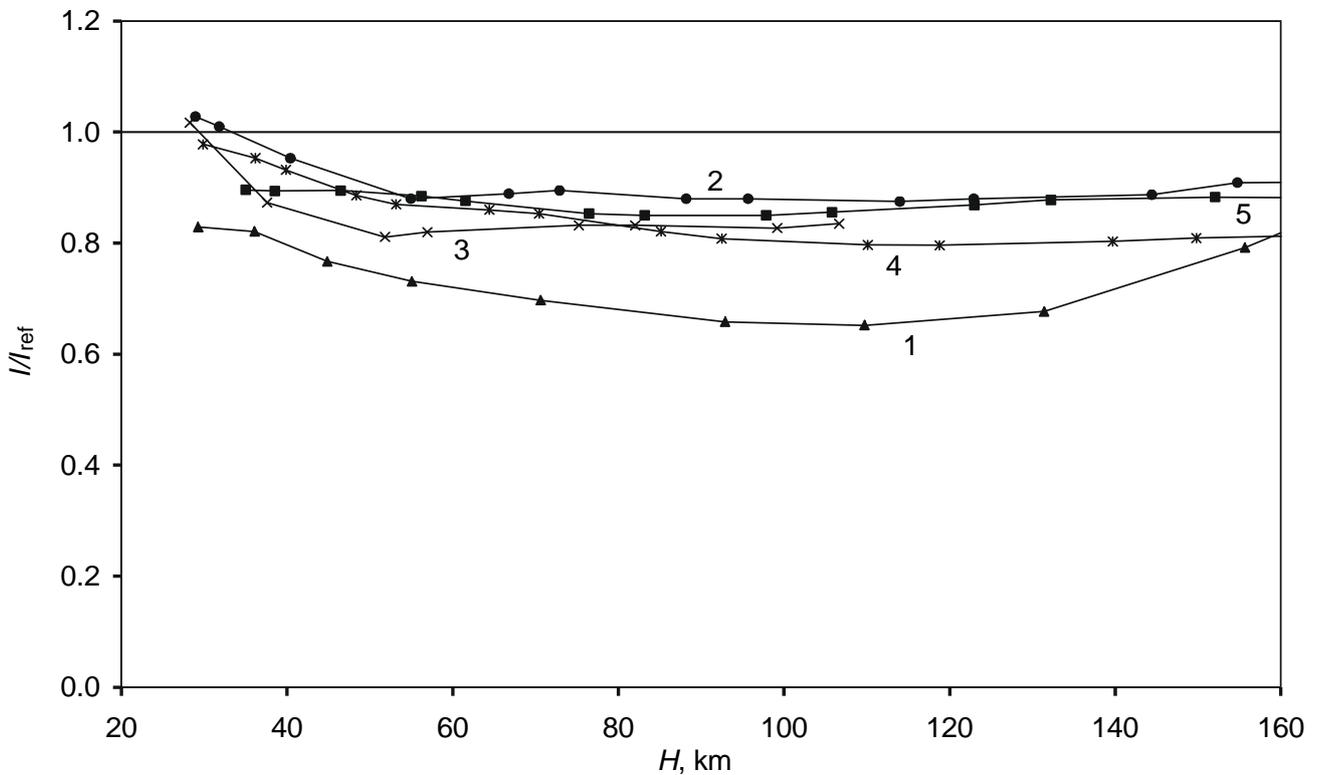

Figure 5. Twilight sky brightness related to the one during reference twilight as the function of effective scattering altitude. Symbols and digital designations are analogous to the Figure 4.

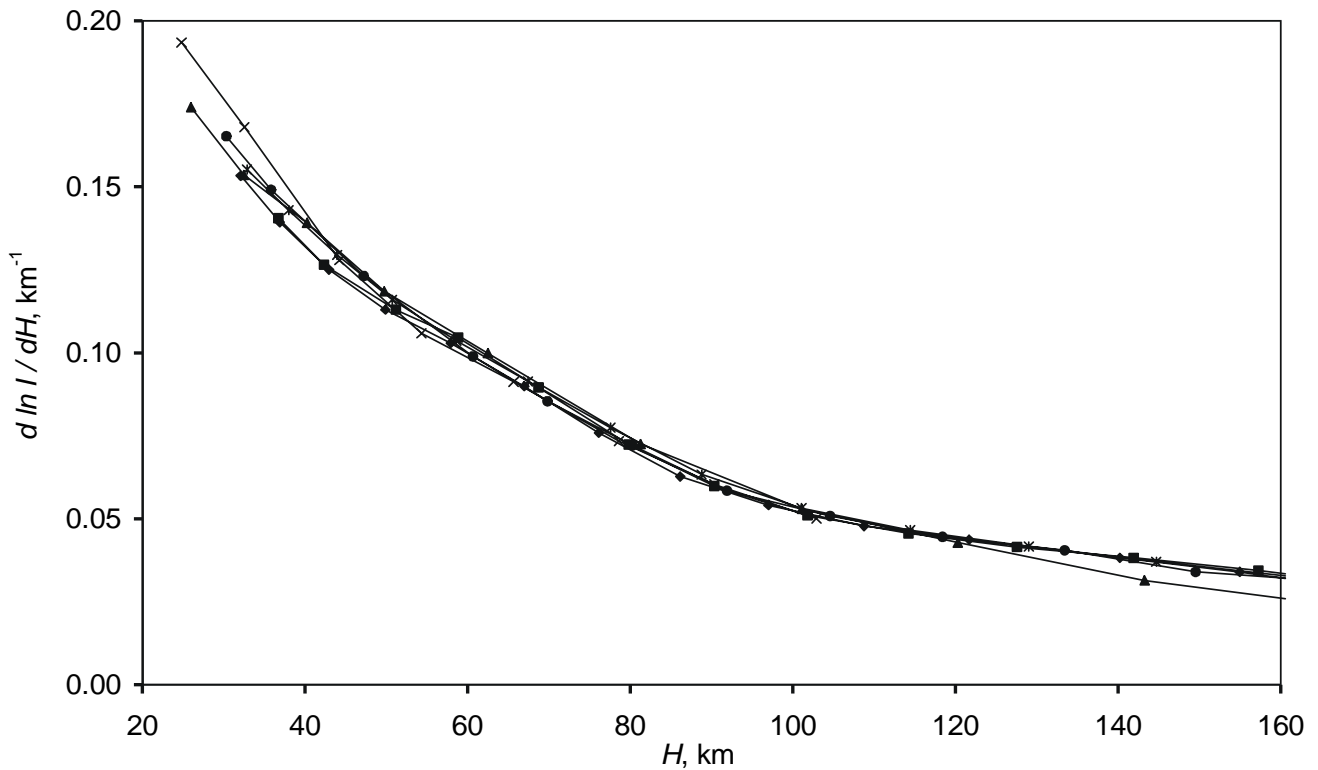

Figure 6. Logarithmic derivative of twilight sky brightness in the zenith by the effective scattering altitude. Symbols correspond to the observation date analogously to the Figure 4.